

6789
\newskip\oneline \oneline=1em plus.3em minus.3em
\newskip\halfline \halfline=.5em plus .15em minus.15em
\newbox\sect
\newcount\eq
\newbox\lett
\newdimen\short
\def\adv{\global\advance\eq by1}
\def\set#1#2{\setbox#1=\hbox{#2}}
\def\nextlet#1{\global\advance\eq
by-1\setbox\lett=\hbox{\rlap#1\phantom{a}}}
\newcount\eqncount\newcount\sectcount\eqncount=0\sectcount=0
\def\sectadv{\global\advance\sectcount by1}
\def\secta{\global\advance\sectcount by1}
\def\equn{\global\advance\eqncount by1\eqno{(\copy\sect.\the\eqncount)} }
\def\put#1{\global\edef#1{(\the\sectcount.\the\eqncount)}           }
\magnification = 1200{\rm}
\voffset 0.2 truein\vsize   8.7 truein\hoffset 0.25 truein
\hsize  6.3 truein\hfuzz 30 pt
\def\mbox#1#2{\vcenter{\hrule \hbox{\vrule height#2in
                \kern#1in \vrule} \hrule}}  
\def\sq{\,\raise.5pt\hbox{$\mbox{.09}{.09}$}\,}
\def\sqb{\,\raise.5pt\hbox{$\overline{\mbox{.09}{.09}}$}\,}
\def\nofirstpagenoten{\nopagenumbers\footline={\ifnum\pageno>0\tenrm
\hss\folio\hss\fi}}
\def\dphi{\dot \Phi}
\def\ddphi{\ddot \Phi}
\def\sig{\sigma}
\def\dsig{\dot \sigma}
\def\ddsig{\ddot \sigma}
\def\dom{\dot \omega}
\def\om{\omega}
\def\ddom{\ddot \omega}
\def\a{\alpha}
\def\rgb{R^2_{GB}}
\def\sign{\rm sign}

\def\np#1#2#3{{\it Nucl. Phys.}\ {\bf B#1}\ ({#2})\ {#3}}
\def\pl#1#2#3{{\it Phys. Lett.}\ {\bf B#1}\ ({#2})\ {#3}}
\def\prl#1#2#3{{\it Phys. Rev. Lett.}\ {\bf #1}\ ({#2})\ {#3}}
\def\prv#1#2#3{{\it Phys. Rev.}\ {\bf D #1}\ ({#2})\ {#3}}

\def\gref#1#2#3#4{{\it #1}\ {\bf #2}\ ({#3})\ {#4}}
\def\cl{\hfil\break}
\def\scs{\ ,\ }

\def\cStrSol{1}
\def\cVen{2}
\def\cSig{3}
\def\cGroSlo{4}
\def\cAntGav{5}
\def\cec{6}
\def\cGBext{7}
\def\cSta{8}
\def\cBra{9}
\def\cOvr{10}
\def\cVil{11}
\magnification=1200
\hsize=6.0 truein
\vsize=8.5 truein
\baselineskip 14pt
\pageno=0
\nofirstpagenoten
\rightline{hep-th/9305025}
\rightline{CPTH-A239.0593}
\rightline{April 1993}
\vskip 1.5truecm
\centerline{\bf {Singularity-free cosmological solutions of the
superstring effective action}}
\vskip 2truecm \centerline{I. Antoniadis,
J. Rizos\footnote{$^*$} {Research
supported by EEC contract Ref. B/SC1\hbox{$*$}-915053.} }
\centerline{Centre de Physique Theorique\footnote{$^{**}$}
{Laboratoire Propre du Centre National de la Recherche Scientifique
UPR A.0014.} } \centerline{Ecole Polytechnique}
\centerline{91128 Palaiseau, FRANCE}
\vskip .3truecm
\centerline{and}
\vskip .3truecm
\centerline{K. Tamvakis}
{\centerline{ Physics Department}
\centerline{University of Ioannina}
\centerline{ GR 45110 Ioannina , GREECE}
\vskip 1truecm \centerline{\bf Abstract}
\par We study the cosmological solutions of the one loop corrected
superstring effective action,  in a Friedmann-Robertson -Walker background,
and in the
presence of
 the dilaton and
modulus fields. A particularly interesting class of solutions is found
 which avoid the initial
singularity  and are consistent with the perturbative treatment of
the effective action.
\vskip .5truecm
\vfill
\eject
\newcount\eqncount
\sectadv
\set\sect{1}
\beginsection{1. Introduction}
\par
\vskip 0.5cm
\par Einstein's theory has been very successful as a classical theory
of gravitational interactions. However, a quantum theory of
gravity would require to incorporate Einstein's theory in a more general
framework. Today our best candidate for such a framework is superstring
theory
which also has the prospect of unification of all other interactions.
Superstrings appear to involve a minimum length which is of the order of
the
Planck scale and, thus, they are expected to lead to drastic modifications
of
the Einstein action at short distances. These modifications can in principle
have cosmological consequences which distinguish string cosmology from
the standard model and provide some indications for a stringy origin of
the universe. Furthermore, one hopes to find resolutions to some puzzles
in Einstein cosmology including the initial singularity problem.

String theory gives rise to two kinds of such modifications. The first is
associated with the contribution of the infinite tower of massive string
modes
and leads to $\alpha'$-corrections, while the second is due to quantum
loop effects. Both these contributions can be studied in the context of
an effective field theory which involves only the massless string modes.
The
effective Langrangian can be derived from string theory using a
perturbative
approach in both the string tension $\alpha'$ and string coupling
expansion. An
alternative direct way to take into account all $\alpha'$-corrections at
the
classical string level is to consider conformal field theories which
describe
exact string solutions in time-dependent gravitational backgrounds
[\cStrSol,\cVen]. However this is not in general sufficient for probing the
short distance behavior of the theory since the gravitational coupling is
dimensionful and quantum corrections become important in this region.

One of the unique properties of the string effective action is that
couplings
are field-dependent. For our purposes, the relevant fields are the dilaton
which plays the role of the string loop expansion parameter, and the moduli
whose vacuum expectation values describe the size and the shape of the
internal compactification manifold. Since these fields have no potential
their
contribution is expected to be important in any cosmological solution of
the
effective action. Here we restrict ourselves to the simple case of a single
modulus field, besides the dilaton, which corresponds to the common
compactification radius.

The tree-level string effective action has been calculated up to several
orders
in the $\alpha'$-expansion in both the sigma model approach [\cSig], where
one considers strings propagating in background fields, and the S-matrix
approach, where the effective action is computed directly from string
scattering amplitudes [\cGroSlo]. It turns out that there is no moduli
dependence of the tree-level couplings. The one loop corrections to the
gravitational couplings have been also calculated recently in the context
of
orbifold compactifications of the heterotic superstring [\cAntGav]. It has
been
shown that there are no moduli dependent corrections to the Einstein term
while
there are non-trivial $R^2$ contributions. They appear as the Gauss-Bonnet
combination multiplied by a function of the modulus field. It is
interesting to
consider the implications of this term to the  the cosmological solutions
of
Einstein equations for two main reasons. Firstly this term is subject to a
non-renormalization theorem which implies that all higher loop moduli
dependent
$R^2$ contributions vanish. On the other hand, it breaks the continuous
isometries of the tree level modulus kinetic terms leaving intact only the
duality symmetries.

The purpose of this paper is to study the evolution of the  equations of
motion of the corresponding effective Lagrangian. We investigate the
asymptotic solutions and we show that the one loop moduli dependent action
contains a class of interesting cosmological solutions which avoid the
initial
singularity. This is possible only for a definite sign of the corresponding
four-dimensional trace anomaly for which the strong energy conditions
[\cec]
related to the modulus energy-momentum tensor can be violated. These
solutions
start from flat space-time in the infinite past, they pass through an
inflationary period and they end up as a slowly expanding universe.
Although in
our analysis we omit terms higher than quadratic in the Riemann tensor, we
will
argue that the above result persists in the full theory under certain
assumptions on the moduli dependence of loop corrections to higher
derivative terms.

The Gauss-Bonnet density multiplied by the dilaton field is already
present at the string tree-level in the next-to-leading order of the
$\alpha'$-expansion and it has been studied extensively in the literature
[\cGBext]. As we will show in Section 3, the addition of this term alone
does
not lead to violation of the energy conditions and thus it cannot provide
any
singularity-free solution, in contrast to the modulus dependent loop
correction.

We also examine the possible existence of time dependent solutions which
fix
asymptotically the vacuum expectation value of the modulus field. This is
motivated from the fact that the Gauss-Bonnet contributions could be viewed
as
a time dependent potential for the modulus with an extremum at the
self-dual point. We show that this possibility can be realized only at
early
times and in singular strong coupling solutions.

This paper is organized as follows. In Section 2  we
review the string effective action and the loop corrections to the
gravitational couplings and we derive the equations of motion for the
coupled
system of graviton, dilaton and modulus field. In Section 3 we study the
cosmological solutions in the simple case where the dilaton is ignored. We
classify all asymptotic solutions and we show by numerical integration that
two
of them can be smoothly joined avoiding the singularity. We justify this
behavior by demonstrating the violation of the weak and strong energy
conditions related to the singularity theorems. Finally, we explore the
parameter space of all initial conditions by the use of the corresponding
phase
diagram. In Section 4 we repeat the same analysis in the presence of the
dilaton field and show that the previous results remain unaffected.  Our
conclusions are summarized in Section 5.

\newcount\eqncount
\vskip 1.0cm
\sectadv
\set\sect{2}
\beginsection{2. The loop corrected effective action and the equations of
motion}
\par
\vskip 0.5cm
\par
Let us consider the universal part of the effective action of any
four-dimensional heterotic superstring model which describes the dynamics
of
graviton, dilaton $S$ and the common modulus field $T$. At the string
tree-level, and up to first order in the $\alpha'$-expansion, it takes the
form
[\cSig,\cGroSlo]:
$$
{\cal L_{\rm eff}} = {1\over 2 \kappa^2} R  + {D S D {\overline S}
\over (S+{\overline S})^2} + 3{D T D {\overline T} \over
(T+{\overline T})^2} + {1\over 8} ({\rm Re}S) R^2_{GB} + {1\over 8} ({\rm
Im}S)
R{\tilde R}\ ,   \equn\put\treel
$$
where $D$ denotes covariant differentiation, $R$ is the scalar curvature,
$\kappa=\sqrt{8 \pi G_N}$ with $G_N$ the Newton's constant, $R^2_{GB}$ is
the
Gauss-Bonnet integrand $$
\rgb=R_{\mu\nu\kappa\lambda}R^{\mu\nu\kappa\lambda} -
4 R_{\mu\nu} R^{\mu\nu}+ R^2\ ,
\equn\put\defgb
$$
and $R{\tilde R} = \eta^{\mu\nu\kappa\lambda}
R_{\mu\nu}{}^{\sigma\tau} R_{\kappa\lambda\sigma\tau}$.\footnote{$^1$}{We
use the conventions $R_{\mu\nu}=R^{\lambda}_{\mu\lambda\nu} \scs
R^{\lambda}_{\mu\nu\kappa}=\partial_\kappa\Gamma^{\lambda}_{\mu\nu}+\dots$,
and
$\eta^{\alpha\beta\gamma\delta}={1\over{\sqrt{-g}}}
\epsilon^{\alpha\beta\gamma\delta}$ with
$\epsilon^{0ijk}=-\epsilon_{ijk}$.}
The inverse of ${\rm
Re}S$ plays the role of the string coupling constant squared while ${\rm
Im}S$
is a pseudoscalar axion. Finally the real part of the complex modulus field
$T$
corresponds to the square of the compactification radius.

At the one loop level the moduli
dependence of the gravitational couplings in the case of the heterotic
string compactified on a symmetric orbifold has been studied in [\cAntGav
]. It
is shown that there are no moduli dependent corrections to the Einstein
term, while the contributions to
the four derivative gravitational terms take the form
$$
\Delta{\cal L}_{\rm eff}= \Delta(T,{\bar T}) R^2_{GB}+\Theta(T,{\bar
T}) R{\tilde R}\ .\equn\put\dl
$$
The moduli dependent functions are defined as
$$
\Delta(T,{\bar T})={{\hat b}_{gr}\over 32 \pi^2} \ln\left[\left(T+{\bar
T}\right) |\eta(i T)|^4\right]\equn\put\dedef
$$
and $\Theta(T,{\bar T})=-i \Delta(T,{\bar T})$ , where
$\eta(\tau)=q^{1/12}\prod_{n\geq1}\left(1-q^{2 n}\right)$, with
$q=e^{i \pi\tau}$, is the Dedekind $\eta$-function. $\Delta(T,{\bar T})$ is
invariant under the duality $SL(2,{\bf Z})$ transformations $T\to 1/T$ and
$T\to
T+i$ which are the discrete subgroup of the continuous $SL(2,{\bf R})$
isometry group of the modulus kinetic terms in {\treel}. The coefficient
${\hat
b}_{gr}$ is proportional to the four dimensional trace anomaly of the N=2
sectors of the theory
$$
{1\over6}(-3 N_V + N_S) - {11\over3}(-3+N_{3/2})\ ,\equn\put\bdef
$$
where $N_S$, $N_V$ and $N_{3/2}$
denote the number of  chiral, vector and spin-3/2 massless supermultiplets.

We now  consider the effective action {\treel}+{\dl} in a spatially flat
homogeneous and isotropic Robertson-Walker background. Since $R {\tilde R}$
vanishes identically in this background, it is consistent with the
equations of
motion to assume for simplicity ${\rm Im}S$ = constant and ${\rm Im}T=0$.
The
latter is required in order to have vanishing derivative of
$\Delta(T,{\bar T})$ with respect to ${\rm Im}T$. Setting the
length scale $\kappa=1$ and defining ${\rm Re}S={1\over g^2}e^\Phi$, where
$g$
is the four-dimensional string coupling, and ${\rm Re}T=e^{2\sig}$ the
effective action takes the form
$$
{\cal S} = \int
d^4x\sqrt{-g}\left[{R\over2}+ {{1\over 4}(D\Phi)^2}+ {{3\over
4}(D\sig)^2} + {1\over 16}(\lambda\ e^\Phi - \delta\ \xi (\sigma))R^2_{GB}
\right] \, \equn\put\action
$$
where $\lambda={2\over g^2}$, $\delta= {{\hat b}_{gr}\over 2
\pi^2}$ and $\xi(\sig)= \ln \left[2 e^{\sig}\eta^4(i e^{\sig})\right]$.

The equations of motion derived from the action {\action} by
varying with respect to the metric, the dilaton and the modulus field
have the form
 $$
T_{\mu\nu}=-{2\over \sqrt{-g}}{{\delta{\cal S}}\over\delta g^{\mu\nu}}=
T_{\mu\nu}^{(1)} + T_{\mu\nu}^{(2)}=0 \ ,
\equn\put\emt $$
$$
{1\over\sqrt{-g}}{\delta{\cal S}\over \delta\sigma}=-{3\over 2} D^{\mu}
D_{\mu}\sigma + {\partial f \over\partial\sigma}R^2_{GB} = 0\ ,
\equn\put\sige
$$
$$
{1\over\sqrt{-g}}{\delta{\cal S}\over \delta\Phi}=-{1\over 2}
D^{\mu} D_{\mu}\Phi + {\partial
f\over\partial\Phi}R^2_{GB} = 0\ ,\equn\put\phie
$$
where
$$\eqalign{
T_{\mu\nu}^{(1)}&=R_{\mu\nu}-{1\over2}g_{\mu\nu}R +
{3\over 2}D_\mu\sig D_\nu\sig-{3\over 4}g_{\mu\nu}
(D \sig)^2
+{1\over 2}D_\mu\Phi D_\nu\Phi-{1\over 4}g_{\mu\nu}
(D \Phi)^2\ , \cr
T_{\mu\nu}^{(2)}&=(g_{\mu\rho}g_{\nu\lambda}+g_{\nu\rho}g_{\mu\lambda})\eta^
{
\kappa\lambda\alpha\beta} D_\gamma\left( {\tilde
R}^{\gamma\rho}{}_{\alpha\beta}D_\kappa f\right) \ ,}
\equn\put\tt
$$
and $f ={1\over 16}(\lambda\ e^\Phi - \delta\ \xi(\sigma))$.

Substituting the spatially flat Robertson-Walker ansatz for the metric
$$
g_{\mu\nu}=(1,-e^{2\omega}\delta_{ij})\equn\put\emet
$$
in {\emt-\tt}
and considering only time dependent fields, we end up with the equations
$$
3\dom^2-3 {\dsig^2\over4}-{\dphi^2\over4}+24{\dot
f}\dom^3=0 \ ,\equn\put\tzz
$$
$$2\ddom +3\dom^2 +3
{\dsig^2\over4}+{\dphi^2\over4} +16 {\dot f}\dom^3 + 8{\ddot f}\dom^2
+16{\dot f}\dom\ddom=0 \ , \equn\put\tii
$$
$$
\ddsig +3 \dom \dsig + \delta {\partial \xi\over
\partial\sigma} \dom^2 (\dom^2+\ddom)=0 \ ,\equn\put\eqsig
$$
$$
\ddphi +3 \dom \dphi - 3\lambda {e^\Phi} \dom^2
(\dom^2+\ddom)=0 \ ,\equn\put\eqphi
$$
where {\tzz} and {\tii} correspond to the $T_{00}$ and $T_{ii}$ components
of
{\emt},  respectively. These two equations are not functionally independent
because of the Bianchi identity related to the conservation of the total
energy
momentum tensor. In fact the linear combination $\dom \times${\tii}$ - \dom
\times ${\tzz}
$-{1\over2}\dsig\times${\eqsig}$-{1\over6}\dphi\times${\eqphi}
yields the time derivative of {\tzz}. Thus, we can reject {\tii} and
consider
the system of {\tzz}, {\eqsig} and {\eqphi} as the independent equations of
motion.

Finally, let us also define here the energy density ($\rho$)
and pressure ($p$) of the dilaton-modulus
matter system. Assuming a perfect fluid form for their energy momentum
tensor
$T_{00} =\rho$, $T_{ii}=p e^{2\om}$, and using {\tzz-\tii,} we get
$$
\eqalign{\rho=&3 \dom^2\cr p=&-(2\ddom+3\dom^2) \ .}\equn\put\defrpb
$$

\newcount\eqncount
\vskip 1.0cm
\sectadv
\set\sect{3}
\beginsection{3. Analysis of the metric-modulus system}
\par
\vskip 0.5 cm
\par
For simplicity we start our analysis by neglecting all the dilaton
related terms. Although $\Phi$ = constant is not a solution of the dilaton
equation of motion {\eqphi}, the metric-modulus system provides a simple
model
to study the effect of the Gauss-Bonnet loop correction
{\dl}. This simplification will be justified in Section 4, where the
full system will be examined, and we will show that the main results of the
following analysis remain valid even in the presence of the dilaton.

The system of equations of motion {\tzz}-{\eqphi} is now reduced to
$$
4\dom^2- {\dsig^2}- 2 \delta{\partial \xi\over
\partial\sigma}{\dsig}\dom^3=0 \ ,\equn\put\ea
$$
$$
\ddsig +3 \dom \dsig +  \delta {\partial
\xi\over \partial\sigma} \dom^2 (\dom^2+\ddom)=0\ .\equn\put\eb
$$
Note that the absolute value of the trace anomaly coefficient
$\delta$ can be absorbed by a time rescaling
$$
t\rightarrow t'= \sqrt{|\delta|}\quad t\equn\put\tresc
$$
which implies that we can replace $\delta$ by its sign. From the expression
{\bdef} one sees that in any theory with $N\le 4$ supersymmetry the sign of
$\delta$ tends to be positive unless there is a considerable excess of
vector
bosons.

The non-linear system {\ea}-{\eb} can be integrated using numerical
methods.
Given initial values for $\sig$, $\om$ and $\dom$, the value of $\dsig$ can
be
determined from {\ea}. Then, the time derivative of {\ea} together with
{\eb}
form a second order system of differential equations which can be
numerically
solved given the above boundary conditions.

Before proceeding to the numerical integration, we shall first derive
analytically the asymptotic solutions in the limits $t\rightarrow\infty$
and
$t\rightarrow 0$. For this purpose we will use the asymptotic expansion of
$\xi'(\sig)\equiv{\partial\xi\over\partial\sig}$ for $\sig\to\pm\infty$:
$$
\xi'(\sigma)\sim -\sign(\sig) {\pi\over3}
e^{|\sig|} \ .\equn\put\aexp
$$
In the limit $t\rightarrow\infty$ we find two kind of solutions. The first
is
obtained using the ansatz
$$
\eqalign{\omega&=\omega_0 + \alpha \ln t \ , \cr\sig&=\sig_0 + \beta \ln t
\ ,
\cr}\equn\put\anz
$$
with $\omega_0$ and $\sig_0$ constants, and it leads to two possibilities:
$$
({\rm A}_\infty): \hskip 4cm \alpha={1\over3}\scs
|\beta|={2\over3} \hskip 4cm\equn\put\sola $$
and
$$
({\rm B}_\infty):\quad|\beta|=2\scs \a^3-\a^2+5\a-1=0\scs
e^{\sign(\beta)\sig_0} = {3(1-\a^2)\over
\delta\pi\a^3}\scs\equn\put\solb
$$
which leads to $\delta >0$ , $\a\sim 0.207$ and ${\rm
sign}(\beta)\sig_0\sim
4.64-\ln\delta$. (A$_\infty$) is actually the asymptotic solution  of the
tree
level system ($\delta =0$), while (B$_\infty$) is a new asymptotic solution
where the Gauss-Bonnet term is important at late times. They both describe
a
slowly expanding universe, while the radius $e^{2\sig}$ is either expanding
or
contracting due to the duality symmetry $\sig\to -\sig$ of equations
{\ea}, {\eb}.

The second kind of solutions is obtained  using the flat space ansatz
$$
\eqalign{\omega&=\omega_0 + \omega_1 t^a \ ,\cr
\sig&=\sig_0 + \beta \ln t \ ,\cr }\equn\put\anzb
$$
which leads to
$$
({\rm C}_\infty):\hskip 1.5cm \a=-1\scs|\beta| = 5\scs
e^{\sign(\beta)\sig_0}
=- {15\over 2  \delta\pi \omega_1^3}\hskip 1.5cm\equn\put\solc
$$

This describes an asymptotically flat universe with slowly expanding (or
contracting) radius.

In the limit $t\to 0$ we find only one asymptotic solution:
$$
({\rm A}_0): \hskip 3.3cm\eqalign{\om=&\om_0+\alpha\ln t\cr
\sig=&\sig_0+\sig_1
t^\beta}\hskip 3.3cm\equn\put\sold
$$
with
$$
\alpha=1\scs\beta=2\scs\sig_1=
{1\over\delta\xi'(\sig_0)}\ . \equn\put\solea
$$
This is a singular solution which also fixes the modulus field to a
constant (but arbitrary) value.

Note that the modulus equation of motion {\eb} for slowly varying $\dom$
can
be considered as describing the motion of $\sig$ in the presence of a
potential
proportional to $\xi (\sig )$. This potential has an extremum at the self-dual
point $\sig=0$ which corresponds to a minimum for $\delta <0$, providing a
possible mechanism to fix the value of the compactification radius.
Unfortunately, the first equation {\ea} requires $\dom =0$ when $\sig=0$
which
leads to vanishing potential for $\sig$. The resulting solution $\sig=0$
and
$\om=$ constant cannot be continuously approached in the asymptotic region.
In
the next Section we will see that in the presence of the dilaton, {\ea} has
a
slowly expanding solution ($\om\sim\ln t$) which leads to a realization of
this mechanism at early times. In addition, the possibility of
fixing the modulus through the same mechanism at finite time, in a region
where
$\dom$ exhibits an extremum, remains open.

Numerical integration of the system {\ea}-{\eb} verifies the existence of
the
above list of asymptotic solutions but it also reveals another very
interesting
characteristic : For $\delta<0$, there exists a region of boundary
conditions, for which the two asymptotic solutions A$_\infty$ and
C$_\infty$ are smoothly joined avoiding the singularity. In fact, starting
from
the asymptotically flat solution C$_\infty$ at the infinite past one is
always
continuously driven to the slowly expanding solution A$_\infty$ at the
infinite future. For the rest of boundary conditions, as well as for
$\delta>0$, the singular solution $({\rm A}_0)$ is recovered. A typical
non-singular solution is presented in figs. 1-2.\footnote{$^2$}{In all
plots
we have used $|\delta|=64$.} Fig. 1 shows that the expansion rate of the
universe $\dom$ starts from a zero value (flat space time) at
$t\to-\infty$, it
grows up to a maximum value, and then it falls down again as $1/t$ at
$t\to\infty$. The scale factor, correspondingly, starts from a constant
value,
it goes through a period of rapid expansion (inflation) and it ends up to a
slowly expanding universe. On the other hand, fig. 2 shows that the modulus
field starts from $-\infty$ corresponding to zero compactification radius
at
the remote past, it passes through the self-dual point $\sig=0$ during the
inflationary period, and it ends up to a slowly expanding regime at the
infinite future\footnote{$^3$}{The duality symmetry implies also the
existence
of the dual solution $\sig\to -\sig$.}. The vanishing of $\sig$ in the
region
where $\dom$ is maximum is a general feature of this kind of solution,
which is
due to the form of the Gauss-Bonnet loop correction as described above.

Figs. 1,2 also show that the obtained non-singular solutions can have all
time
derivatives of $\omega$ and $\sig$ less than one in Planck units which is
consistent with our approximation of neglecting higher derivative terms in
the
effective action. This is in general sufficient provided that the
moduli-dependent coefficients of the higher order terms are not large
enough to
compensate the derivative suppression. Under this assumption the main
features
of these solutions are expected to survive after higher order corrections
are
taken into account.

The avoidance of the singularity  is accompanied, as expected, by a
violation
of the weak and strong energy conditions [\cec] related to the
energy-momentum
tensor of the modulus field, which are illustrated in fig. 3. This
violation
can be demonstrated analytically in the following way. The system of
equations
consisting of the time derivative of {\ea} together with {\eb} can be
solved
for $\ddom$ and then ${\xi'(\sigma)}$ terms can be removed using {\ea}.
Using
{\defrpb} we obtain
$$
{\eqalign{\rho+p=& - 2\ddom= 2 \dom^2
{{16\dom^4+24 \dom^2\dsig^2+5\dsig^4 - 4
\delta\xi''(\sig)\dom^2\dsig^4}\over{16\dom^4-8\dom^2\dsig^2+5 \dsig^4}}\ ,
\cr
\rho+3p=& -6(\dom^2+\ddom)= 24\dom^4\dsig^2 {{8 - \delta\xi''(\sig)
\dsig^2}\over{16\dom^4-8\dom^2\dsig^2+5 \dsig^4}}}}\equn\put\ec
$$
where
$\xi''(\sig)\equiv{\partial^2\xi\over\partial^2\sig}$. It is easy to see
that since $\xi''(\sig)<0$, equation {\ec} implies that $\rho+p>0$ and
$\rho+3p>0$ for $\delta>0$. Thus, the energy conditions can never be
violated
in this case and we cannot avoid the initial singularity. This is also the
result of the numerical integration, for $\delta>0$, which always
leads to the singular solution A$_0$ at early times.

On the contrary, as seen from {\ec}, $\delta<0$ allows for solutions
which explicitly violate  both energy conditions. Of course this is in
general
a necessary but not sufficient requirement in order to avoid singularities.
In
our case, the numerical integration of the system {\ea}-{\eb} has shown
that
the violation of the energy conditions at some instant is also sufficient.
In
fact, the non-singular solutions can be obtained by imposing
$\ddom\sim\rho+p=0$, which implies $\rho+3p<0$, at the starting point of
our
integration. The initial values for $\sig$ and $\dom$ at this point obey a
constraint which can be derived using equations {\ec} and {\ea}. For a
specific
value of the modulus $\sig$ the expansion rate $\dom$ is given by
$$
\eqalign{4 \nu^4 z^5 -5 \nu^4 z^4+40 \nu^2&
z^3 -8(2+7 \nu^2) z^2 + 96 z -144=0}\scs
z>{5\over4}\equn\put\eee
$$
where $z\equiv \delta \xi''(\sig)\dom^2_{\rm max}$ and
$\nu\equiv{\xi'(\sig)\over\xi''(\sig)}$ satisfying $-1<\nu<1$ due to the
properties of $\xi(\sig)$. One can show that the above equation has at
least
one solution for $z$ for any value of $\nu$. In terms of our original
variables $\sig$ and $\dom$, these solutions are plotted in fig. 4.

The entire region of initial conditions which lead to singular or
non-singular
solutions for $\delta <0$ can be explored by the use of the corresponding
phase
diagram. In fact, the system of second order differential equations
{\ea-\eb}
can be reduced to a single first order equation for the variable $\dom$ as
a
function of $\sig$. First {\ea} can be solved for $\dsig$ in terms of
$\dom$
and $\sig$:
$$
\dsig=-\delta\xi'(\sig) \dom^3 \pm |\dom|\sqrt{\delta^2\xi'(\sig)^2 \dom^4
+4}\
. \equn\put\eqy
$$
On the other hand, the time-derivative of {\ea} together with {\eb} leads
to
{\ec} which can written in the form:
$$
{d{\dom}\over d\sig}=-{\dom^2\over\dsig}
{{16\dom^4+24 \dom^2\dsig^2+5\dsig^4 - 4\delta
\xi''(\sig)\dom^2\dsig^4}\over{16\dom^4-8\dom^2\dsig^2 + 5\dsig^4}}\ .
\equn\put\eqx
$$
Inspection of {\eqy} shows that the two branches defined by the two signs
of
the square root correspond to two disconnected classes of solutions
associated
with positive or negative values of $\dsig$. These are related by a duality
transformation $\sig\to -\sig$, and thus we can restrict to the case
$\dsig >0$.

The phase diagram of {\eqx} is presented in fig. 5. The asymptotic
solutions C$_{\infty}$ and A$_{\infty}$ are located in the regions $\sig\to
-\infty$ and $\sig\to +\infty$, respectively, while the singular A$_0$
appear
in the region $\dom\to \infty$, $\sig >0$. As one sees, for negative
initial
values of $\sig$ non-singular solutions are obtained for all values of
$\dom$. They always interpolate between an asymptotically flat and a slowly
expanding universe. The maxima of $\dom$ correspond to points of the curve
($\dsig >0$) shown in fig. 4. On the other hand, for positive initial
values of
$\sig$ singular solutions are obtained for all values of $\dom$ lying above
a
critical curve (bold line). Note that they never end up to asymptotically
flat
space time. Comparing to other known non-singular solutions [\cSta,\cBra]
derived from higher derivative effective gravity theories, our solutions
extend
over an infinite region of the phase space where the maximum value of
curvature
is not bounded.

As already mentioned, another interesting characteristic of these
non-singular
solutions is that they contain an inflationary regime.\footnote{$^4$}{A
discussion of the contribution of the string loop moduli dependent
corrections
to the gravitational couplings in connection with the inflationary
solutions
of ref. [\cSta] was reported in ref. [\cOvr].} This is expected at least
during
the period when the energy conditions are violated due to the development
of
negative pressure. The numerical analysis shows that the amount of
inflation
increases with rising values of $\delta^{1/2}\dom_{\rm max}$ given in fig.
4.
For values of order one only a few e-foldings are obtained (see fig. 1), while
larger values of this parameter break the validity of the perturbative
treatment of the effective action. In any case, these solutions provide an
example of a cosmological model which is driven to an inflationary era and
then
exits in a finite time. This is in contrast to the behavior of other
non-singular solutions which start from de Sitter space in the remote past
[\cSta,\cBra]. In fact the solutions presented here start asymptotically
from
flat space-time which is welcome as it has been recently argued that
eternal
de-Sitter inflation may not be possible without a beginning [\cVil].
Moreover,
their characteristics are very similar to those arising in the context of a
``pre-big-bang" scenario motivated by generalized scale-factor duality
symmetries of string theory [\cVen].

The properties of the non-singular solutions depend crucially
on the form of the modulus dependent function $\xi(\sig)$ arising from the
string loop corrections to the dimensionless gravitational couplings
{\action}.
For instance if $\xi(\sig)\sim\sig^2$ de Sitter space is obtained
asymptotically. On the other hand if one considers the dilaton field
instead
of the modulus in the presence of the first order in $\alpha'$ Gauss-Bonnet
interaction {\action}, the singularity can never be avoided. In fact the
analysis is equivalent to that of the metric-modulus system {\ea}-{\eb}
with
the substitutions $\sig\to{\Phi\over 3}$, $\delta\xi'(\sig)\to -\lambda
e^{\Phi}$. Then, the energy conditions {\ec} have also the same form with
$-\delta\xi''(\sig)$ replaced by $\lambda e^\Phi$ and they cannot be
violated since $\lambda=2/g^2$ is positive.

\newcount\eqncount
\vskip 1.0cm
\sectadv
\set\sect{4}
\beginsection{4. Analysis including the dilaton terms}
\par
\vskip 0.5cm
\par
We now extend our analysis including the dilaton contributions in the
effective action {\action}, which leads to the equations of motion {\tzz},
{\eqsig} and {\eqphi}. A time dilatation combined with a shift in the
dilaton field,
$$
\eqalign{t\rightarrow t'&=
\sqrt{|\delta|}\quad t\cr \Phi\rightarrow
\Phi'&=\Phi+2\ln|\delta|-\ln\lambda}\equn\put\trescb
$$
can be used to eliminate both $\lambda$ and the absolute value of
${\delta}$.
In analogy with the analysis of the previous Section we shall first derive
the
asymptotic solutions. As one can see from {\aexp}, the asymptotic behavior
of
the modulus dependent Gauss-Bonnet coefficient $\xi(\sig)$ for large $\sig$
is
proportional to $e^{|\sig|}$, while the dilaton dependent coefficient is
proportional to $e^{\Phi}$. It follows that the asymptotic solutions of the
metric-modulus system derived in Section 3 will survive in this case, with
the
dilaton either being negligible or behaving similarly to the modulus. In
addition, some new solutions are expected for dominant large asymptotic
values
of the dilaton field.

In the limit $t\rightarrow\infty$ the asymptotic solutions (A$_\infty$) and
(B$_\infty$) of Section 3 are extended with the dilaton being of the form
$\Phi=\phi_0 + \gamma \ln t$ to the solutions:
$$
({\rm A}_\infty'): \alpha={1\over3}\scs 9 \beta^2+ 3 \gamma^2
=4\equn\put\solad
$$
and
$$
({\rm B}_\infty'): {|\beta|=\gamma =2\scs\a\sim
0.205\scs {\rm sign}(\beta)\sig_0 \sim
4.67-\ln\delta\scs\delta>0\scs\phi_0\sim3.62\ .} \equn\put\solbd
$$
Furthermore, (A$_\infty$) can also be extended with the dilaton of the
form $\Phi=\phi_0+\phi_1 t^\gamma$:
$$
({\rm A}_\infty''):\hskip 2cm
\alpha={1\over3}\scs|\beta|={2\over3}\scs\gamma=-2\scs
\phi_1=-{e^{\phi_0}\over 54} \ . \hskip 2cm\equn\put\solbdd $$
In the above solutions the dilaton field either grows to
plus or minus infinity (A$_\infty'$ or A$_\infty'$ and B$_\infty'$)
corresponding to weak or strong string coupling, respectively, or it
reaches
asymptotically an arbitrary constant value (A$_\infty''$). As in the
metric-modulus case in the solution A$_\infty'$ the Gauss-Bonnet terms
become
irrelevant at large times.

Similarly, the asymptotically flat solution C$_\infty$ can be extended with
the dilaton being either of the form $\Phi=\phi_0 + \phi_1 t^\gamma$:
$$
({\rm C}_\infty'):\hskip .5cm {\a=-1\scs|\beta|=-\gamma=5\scs
\phi_1={1\over5}e^{\phi_0}\om_1^3\scs
e^{{\sign(\beta)}\sig_0} = -{15\over2 \pi \om_1^3\delta}}\ ,\hskip
.5cm\equn\put\solbc  $$
or of the form $\Phi=\phi_0 +\gamma \ln t$:
$$
 ({\rm C}_\infty''):\a=-1\scs|\beta|=\gamma =5\scs
\delta>0 \scs e^{{\sign(\beta)}\sig_0}= -{15\over2 \pi \om_1^3\delta}\scs
e^{\phi_0}=-{5\over6\om_1^3}\ .\equn\put\solbd
$$
In C$_\infty'$ the dilaton goes to a constant, while C$_\infty''$ is a weak
coupling solution.

There are also three new solutions where the modulus field goes
asymptotically
to an arbitrary constant. They are obtained from A$_\infty''$ (D$_\infty$,
D$_\infty'$) or C$_\infty'$ (E$_\infty$) by interchanging the role of
$\sig$
and $\Phi$, $\sig=\sig_0+\sig_1 t^\beta$ and $\Phi=\phi_0+\gamma \ln t$ :
$$
({\rm D}_\infty):\hskip 1.3cm \a={1\over3}\scs
|\gamma|={2\over\sqrt{3}}\scs\beta=-2\scs
 \sig_1={1\over162}\delta\xi'(\sig_0)\ ,\hskip 1.3cm\equn\put\solbz
$$
$$
({\rm D}_\infty'):\hskip .4cm \a\sim0.223\scs
\gamma=2\scs\beta=-2\scs
 \sig_1\sim0.002\delta\xi'(\sig_0)\scs\phi_0\sim3.24\ ,\hskip
.4cm\equn\put\solbzp $$
$$
({\rm E}_\infty):\hskip .8cm\a=-1\scs-\beta=\gamma =5\scs
e^{\phi_0}= -{5 \over 6 \om_1^3}\scs
\xi'(\sig_0)=-{15 \sig_1\over\delta\om_1^3}\ ,\hskip .7cm\equn\put\solbe
$$
with $\sig_0\ne0$.

The same procedure can be followed to derive the singular solutions in the
limit $t\to 0$. The solution A$_0$ is extended with the dilaton field going
to a constant, $\Phi=\phi_0+\phi_1 t^\gamma$
$$
(A_0') :{\eqalign{\alpha=1\scs&\beta=\gamma=2\cr
\sig_1= {{\delta} \xi'(\sig_0)\over \delta^2\xi'(\sig_0)^2+3 e^{2 \phi_0}}
\scs& \phi_1=- {3 e^{\phi_0}\over \delta^2\xi'(\sig_0)^2+3 e^{2
\phi_0}}}}\equn\put\pardh
$$
Furthermore, for the same form of $\om$ and $\delta<0$, two new
solutions can be obtained with the modulus behaving logarithmically
$\sig=\sig_0 +\beta \ln t$  and the dilaton being either of the form
$\Phi=\phi_0 + \gamma \ln t$
$$
({\rm B}_0):\hskip 1.5cm \alpha=1\scs
|\beta|=-\gamma\scs e^{\phi_0}=-{\delta\pi\over3\delta}
e^{-{\sign(\beta)}\sig_0}\ ,\hskip 1.5cm\equn\put\eeb
$$
or of the form $\Phi=\phi_1 t^\gamma$
$$
({\rm C}_0) :\hskip .5cm \alpha=1\scs |\beta|=2, \gamma=-2, \phi_1=
-\sqrt{-\pi\delta}\exp({{-{\sign(\beta)}\sig_0\over2}})\ .\hskip .5cm
\equn\put\pardg $$
B$_0$ is a weak coupling solution, while C$_0$ is a strong coupling one.

Finally, for the same logarithmic behavior of $\om$ and $\delta>0$
there is one more singular solution
$$
({\rm D}_0): \hskip 3cm\eqalign{\Phi&=\phi_0 + \gamma \ln t \ ,\cr
\sig&=\sig_0 \exp{-\beta\over t} \,\hskip 3cm }\equn\put\anzbe
$$
with
$$
{\eqalign{\gamma=2\scs
3\a^3-3\a^2+5\a-1=0&\scs\beta=\delta \xi''(0)\alpha^3(\alpha-1)>0\cr
\exp(\phi_0)={2(3\a-1)\over 3(1-\a)\a^3}\ ,}}\equn\put\soldg
$$
which leads to $\a\sim0.223$, $\beta\sim0.019 \delta$, $\phi_0\sim3.24$.
This is
a strong coupling asymptotic solution where the modulus field
approaches the self-dual point in a non-analytic way. It provides an
example
of realizing the mechanism described in Section 3 according to which the
Gauss-Bonnet term in {\eqsig} acts as a potential for the modulus with a
minimum at the self-dual point.

The integration of the non-linear system {\tzz, \eqsig, \eqphi} can be
performed numerically following a similar procedure we used in the previous
Section for the metric-modulus case. However, in the presence of the
dilaton the
phase space is enlarged considerably since two more initial values are
required for $\Phi$ and $\dphi$ in addition to $\sig$, $\om$ and $\dom$.
Instead of presenting a detailed numerical investigation which will not be
very
illuminating, we concentrate to the analysis of some physically
interesting cases. In the case $\delta>0$, we verified that the fixed
modulus
singular solution D$_0$ {\soldg} is obtained for a large region of the
parameter space. It turns out that the modulus field approaches at $t=0$
its
self-dual point by dumping oscillations consistently with the effective
potential interpretation (see fig. 6).

The case $\delta<0$ is more interesting since it admits non-singular
solutions. They could be in principle derived by the method described in
Section 3 which consists of starting the numerical integration at a point
where the energy conditions are violated, for instance when $\ddom=0$. It
turns
out that this method is not very useful because, in this case, the
violation of
energy conditions although necessary is not sufficient to avoid the initial
singularity [\cec]. An alternative way would be to start the integration
with
an asymptotically flat solution in the infinite past, which in the
metric-modulus case was shown to lead always to non-singular solutions.
This
turns out to be the case even in the presence of the dilaton. Starting from
C$_\infty'$, which extends C$_\infty$ with a negligible dilaton, one is
smoothly
driven to A$_\infty'$ in the infinite future. This confirms that the main
characteristics of such solutions are a consequence of the modulus
dependent
string loop correction to the Gauss-Bonnet term and they do not depend on
the
existence of the dilaton. In a typical solution of this kind the
scale factor and the internal radius behave similarly to those
obtained in the absence of the dilaton (figs. 1,2). The dilaton evolution
is
presented in fig. 7; it starts from a constant value in the remote past and
grows logarithmically towards strong coupling in the future. During the
inflationary period when the modulus passes through its self-dual point,
the
dilaton also jumps to its maximum value. This can be easily understood by
inspection of its equation of motion {\eqphi}, where the Gauss-Bonnet term
plays the role of a runaway potential for $\dom\sim$ constant. In the limit
$t\to\infty$, although a weak coupling solution of the form A$_\infty'$ or
asymptotically constant dilaton of the form A$_\infty''$ are not a priori
excluded, strong coupling seems to be preferred at least in the restricted
region of the parameter space we scanned. In any case, at late times the
effective action {\action} should probably be modified by the addition of a
dilaton and modulus potential arising from supersymmetry breaking or other
non
perturbative effects, which would stabilize these fields.

\vskip 1.0cm
\noindent{\bf 5. Conclusions}\cl
\vskip 0.5 cm
\par
In this work we have examined the cosmological implications of the moduli
dependent loop corrections to the gravitational couplings of the
superstring
effective action in the case of orbifold compactifications. These
corrections
consist of the Gauss-Bonnet integrand multiplied by a universal non-trivial
function of the moduli fields and a numerical coefficient $\delta$ which
depends on the massless spectrum of every particular model. We first
derived the
equations of motion for the metric, the dilaton and the modulus
corresponding
to the common compactification radius, and we classified all asymptotic
solutions. Among them, there is one where the internal radius is fixed at
its
self-dual point as the universe approaches the initial singularity.

In the case of negative sign for the parameter $\delta$, we have shown that
the strong energy condition $\rho +p>0$ associated to the stress energy
tensor
of the modulus can be violated, leading to an inflationary period and
providing
the possibility of avoiding the initial singularity. In fact a numerical
analysis of the system has verified the existence of a class of
non-singular
solutions which interpolate between an asymptotically flat and a slowly
expanding universe with a period of rapid expansion, when the modulus
field passes through its self-dual value. This is in contrast to the
behavior
of other non-singular solutions which have been proposed in the literature,
where de Sitter space was always obtained at ``early" times [\cSta,\cBra],
but
it shares the properties of a ``pre-big-bang" phase proposed in ref.
[\cVen].

Our solutions depend crucially on the form of the modulus dependent string
loop
corrections, while the dilaton contribution is negligible and can be
ignored.
Furthermore all time derivatives can remain bounded in the perturbative regime,
which is consistent with our approximation of neglecting higher derivative
terms in the effective action. Finally, the whole parameter range has been
explored in the absence of the dilaton by the study of the corresponding
phase
diagram and we found that the class of non-singular solutions extends over
an infinite region of the phase space.

\vskip 1.0cm
\noindent{\it Acknowledgments}\cl
\vskip.5cm
This work was supported in part by EEC contracts SC1-0394-C, SC1-915053 and
SC1-CT92-0792. One of us (K. T.) thanks the Centre de Physique Th\'eorique
of
the Ecole Polytechnique for its hospitality during completion of this
work.

\vfill
\eject
\centerline{\bf References}
\vskip 1cm
\item{\cStrSol.}
R. C. Myers, \pl{199}{1987}{37}; I. Antoniadis, C. Bachas, J.
Ellis and D. V. Nanopoulos, \pl{211}{1988}{393}; \np{328}{1989}{117};
M. M\"uller, \np{337}{1990}{37}; A. A. Tseytlin and C. Vafa,
\np{372}{1992}{443}; C. Kounnas and D. L\"ust, \pl{289}{1992}{56}; A. A.
Tseytlin, DAMTP-92-15-REV (1992).\cl
\item{\cVen.} M. Gasperini, J. Maharana and G. Veneziano,
\pl{296}{1992}{51};
M. Gasperini and G. Veneziano, preprint CERN-TH.6572/92.\cl
\item{\cSig.}
C. G. Callan, D. Friedan, E. J. Martinec
and M. J. Perry, \np{262}{1985}{593}; \np{278}{1986}{78};  E. S. Fradkin
and A.
A. Tseytlin, \pl{158}{1985}{316}; \np{262}{1985}{1}; A. Sen,
\prv{32}{1985}{2102}; \prl{55}{1985}{1846}.\cl
\item{\cGroSlo.}
D. J. Gross and J. H. Sloan, \np{291}{1987}{41}.\cl
\item{\cAntGav.}
I. Antoniadis, E. Gava and K. S. Narain, \pl{283}{1992}{209};
\np{393}{1992}{93}.\cl
\item{\cec.}
R. Penrose, \prl{14}{1965}{57}; S. Hawking, \gref{Proc. R. Soc.
London}{A300}
{1967}{182}; S. Hawking and R. Penrose, \gref{Proc. R. Soc.
London}{A314}{1970}{529}.\cl
\item{\cGBext.}
D. G. Boulware and S.
Deser,\prl{218}{1985}{2656};\pl{175}{1986}{409}; J. T. Wheeler,
\np{268}{1986}{737}; S. Kalara, C. Kounnas and K.Olive,\pl{215}{1988}{265};
 C. Wetterich, \np{324}{1989}{141}; A. A. Tseytlin, Proceedings of
the International Workshop ``String Quantum Gravity and Physics at the
Planck
Energy Scale", Erice, June 1992, and references therein. \cl
\item{\cSta.}
A. A. Starobinsky, \pl{91}{1980}{99}.\cl
\item{\cBra.}
V. Mukhanov and R. Bradenberger,
\prl{68}{1992}{1969}; R. Branderberger, V. Mukhanov and A. Sornborger,
preprint BROWN-HET-891,(1993).\cl
\item{\cOvr.} B. A. Ovrut, talk presented at the 26th
Workshop ``From Superstrings to Supergravity", Erice, December 1992.\cl
\item{\cVil.} A. Vilenkin, \prl{46}{1992}{2355}.\cl

\vfill\eject
\baselineskip=20 pt

\noindent{\bf Figure captions\cl}
{\bf Fig. 1}. The scale factor $e^\om$ (continuous line) and
the Huble expansion rate $\dom$ (dashed line) for the non-singular solution
($\delta<0$).\cl
{\bf Fig. 2}. The modulus field $\sig$ (continuous line) and
its derivative $\dsig$ (dashed line) for the  non-singular solution.\cl
{\bf Fig. 3}. $\rho+p$ (dashed line) and $\rho+3p$ (continuous line) for
the
non-singular solution.\cl
{\bf Fig. 4}. The maximum of the expansion rate $\dom_{max}$ as a function
of
the modulus $\sig$ for $\dsig>0$ (continuous line) and $\dsig<0$ (dashed
line).\cl
{\bf Fig. 5}. The phase diagram of $\dom$ as a function of $\sig$;
continuous
lines correspond to non-singular solutions and dashed lines to singular
ones.
The boundary of the two regions is plotted by a bold line.\cl
{\bf Fig. 6}. The scale factor $e^\om$ (dashed line) and the modulus
$\sig$ (continuous line) in a singular solution ($\delta>0$) where the
modulus
approaches its self-dual point ($\sig=0$) at early times.\cl
{\bf Fig. 7}. The dilaton field $\Phi$ (continuous line) and its derivative
$\dphi$ (dashed line) for the  non-singular solution.\cl

\vfil
\end